# Compressed Sensing on the Image of Bilinear Maps

Philipp Walk
Lehrstuhl für Theoretische Informationstechnik
Technische Universität München
Theresienstrasse 90, 80290 München
Email: philipp.walk@tum.de

Peter Jung
Lehrstuhl für Informationstheorie und
Theoretische Informationstechnik
Technische Universität Berlin
Einsteinufer 25, 10587 Berlin
Email: peter.jung@mk.tu-berlin.de

*Abstract*—For several communication models, the dispersive part of a communication channel is described by a bilinear operation $T$ between the possible sets of input signals and channel parameters. The received channel output has then to be identified from the image $T(X,Y)$ of the input signal difference sets $X$ and the channel state sets $Y$. The main goal in this contribution is to characterize the compressibility of $T(X,Y)$ with respect to an ambient dimension $N$. In this paper we show that a restricted norm multiplicativity of $T$ on all canonical subspaces $X$ and $Y$ with dimension $S$ resp. $F$ is sufficient for the reconstruction of output signals with an overwhelming probability from $\mathcal{O}((S+F)\log N)$ random sub-Gaussian measurements. Thus, in this case, the number of degrees of freedom of each output grows only additively instead of multiplicatively with the input dimensions (sparsity) $S$ and $F$. This is a relevant improvement in the output compressibility and suggests a substantially reduced rate in compressed sampling algorithms.

## I. INTRODUCTION

From an abstract point of view, a dispersive communication channel is given as a bilinear operation $T(\mathbf{s}, \mathbf{h})$ between the input signal $\mathbf{s}$ and the channel parameter $\mathbf{h}$ plus additive noise $\mathbf{n}$. For example, on $\mathbb{R}^N$ this action could be represented as a matrix multiplication, i.e. the received signal is then:

$$\mathbf{r} = \mathbf{H}\mathbf{s} + \mathbf{n}. \qquad (1)$$

and the corresponding channel matrix $\mathbf{H} \in \mathbb{R}^{N \times N}$ is given by the action $\mathbf{Hs} = T(\mathbf{s}, \mathbf{h})$. If the channel matrix is known at the receiver and the possible input signals $\mathbf{s}$ exhibit a linear structure the set of all outputs $\mathbf{Hs}$ is a linear space. However, if the channel matrix has unknown parameters and the signal set is for example only a union of subspaces, the set of all possible outputs usually looses its linearity and the receiver is thus confronted with the determination of $\mathbf{r}$ from a non-linear manifold. Then, in order to provide efficient sampling and reception in such a non–coherent setting, it is of fundamental importance to characterize the complexity of the output set and relate it to structural properties of the channel and of the possible transmit signals. In compressed sensing for example, such an assumption is the sparsity $S$ of the signal set, i.e. the transmitter might operate in a peaky fashion and the data is concentrated only on unknown but small sets of $S \ll N$ sample points. Let us denote this union of so called canonical subspaces with $\Sigma_S$. Even more, the set of non–zero coefficients in the vector $\mathbf{h} \in \mathbb{R}^N$ during transmission can be small as well, i.e. concentrated on subsets of cardinality $F \ll N$ such that $\mathbf{h} \in \Sigma_F$. The intrinsic dimension for sampling the signal $\mathbf{r}$ in (1) is $S + F$ as soon the support of signal and channel coefficients are known to the transmitter, whereby the image of this particular subspaces under $T$ is contained in some subspace up to dimension $SF$. In both cases an additional factor, logarithmic $N$, is necessary for the unknown location of the supports. However, in the worst case, the sparsity of the inputs to $T$ behave *multiplicatively*. The essence of this contribution is to establish conditions on $T$ under which the overall compressibility still behaves only *additively* in $S$ and $F$. This has a relevant impact on the complexity of the output set $T(\Sigma_S, \Sigma_F)$ and hence substantially improve the rate in compressed sampling. Furthermore, understanding the coupling between two sparse sources directly generalizes to the coupling of finitely many sources.

In their recent work [1], Hedge and Baraniuk considered $(S,F)$-sparse circular convolutions in $\mathbb{R}^N$, see Section III, and formulated a *restricted isometry property* (RIP) for the set of all differences in the union of the image of all $(S,F)$-sparse circular convolutions. It turns out that their proof approach leads to difficult mathematical problems, which according to the authors' knowledge are still unsolved. Since the proof in [2, Lemma 5.1] relies on a linear structure, which may not be present for the image of bilinear maps, more strict conditions on $T$ and the input sets $X$ and $Y$ are needed to control the norm of the output set. However, the authors could establish in this work the result of Hedge and Baraniuk for a certain restriction on Euclidean convex cones. Another special case occurs for $S$ and $F$ dimensional subspaces, which are "properly" separated. In this case, the image under the circular convolution is isomorphic to the set of all simple tensors, which in turn is isomorphic to the set of rank-1 operators from $X$ to $Y$. Hence the main theorem implies certain results from the theory of rank-1 matrix recovery [3], [4], [5]. Moreover, the developed framework applies to all bilinear operations which have a certain *restricted norm multiplicativity property* on convex cones in an arbitrary basis. This enables compressed sensing on "sparse" output sets, which can not be written as a finite union of subspaces, and leads to *generalized structured sparsity models* [6], [7], [8].

The outline of this paper is as follow. In Section II the fundamental concept of bilinear maps $T$ on finite dimensional Euclidean spaces is introduced. The authors formulate a sufficient condition in Definition 1, which ensures a better probability as in [2] for the RIP on the image of such bilinear maps. Some important and simple couplings $T$ for certain communication scenarios are discussed in Section III. Here the main result is applied to some channel models, e.g. circular convolutions, which establishes an additive behavior of the sparsity $S$ and $F$ of the signal inputs and channel states. Section IV concludes this work with a conjecture for the RIP on the image of all $(S,F)$-sparse circular convolutions.

## II. MAIN RESULT

Every bilinear map $T\colon \mathbb{R}^N \times \mathbb{R}^N \to \mathbb{R}^N$ is a binary operation on $\mathbb{R}^N$ and defines therefore a multiplication on $\mathbb{R}^N$. The $N$ dimensional linear space $\langle \mathbb{R}^N, +, \cdot, T\rangle$ is then called an *associative algebra*[1] *over* $\mathbb{R}$. Since we are interested in a stable embedding of the output (image), we have to equip the algebra with a norm $\|\cdot\|$. In our main Theorem 1 we need for the proof technique, based on [2], a nesting in the $\ell^2$-norm, i.e. we have to bound the norm of the output $\mathbf{z}$ by the product of the norms of the inputs $\mathbf{x}, \mathbf{y}$ from below and above. This is a very strict property for algebras since this would imply that the nullspace of $T$ is $\mathcal{N} := (\mathbf{0}, X) \cup (Y, \mathbf{0})$, i.e. $T$ is *non-singular* on $X \times Y$. If we remove all singularity points in $X \times Y$ we obtain a subset $O \subset X \times Y$ on which $T$ is non-singular. But then there could still exist a sequence $\mathbf{o}_n \in O\setminus\mathcal{N}$ such that $\lim_n T(\mathbf{o}_n) = \mathbf{0}$. The following definition exclude such sequences and provides how close $\mathbf{0}$ can be approached.

*Definition 1 (Restricted norm multiplicativity property):* Let $X, Y \subset \mathbb{R}^N$. Then the bilinear map $T\colon X \times Y \to \mathbb{R}^N$ has the *restricted norm multiplicativity property* (RNMP), if

$$0 < \alpha := \sup_{\substack{O \subset X \times Y \\ T(O) = T(X,Y)}} \inf_{(\mathbf{x},\mathbf{y}) \in O\setminus\mathcal{N}} \frac{\|T(\mathbf{x},\mathbf{y})\|}{\|\mathbf{x}\|\|\mathbf{y}\|}. \tag{2}$$

Moreover, we define the universal upper bound by

$$\beta := \inf_{(\mathbf{x},\mathbf{y}) \in X \times Y \setminus \mathcal{N}} \frac{\|T(\mathbf{x},\mathbf{y})\|}{\|\mathbf{x}\|\|\mathbf{y}\|}. \tag{3}$$

*Remark 1:* It is known [9], that for finite–dimensional algebras there always exists $\beta < \infty$. Obviously, $\beta$ is always an upper bound on $O$ and this simplification will become relevant in the proof of Theorem 1. If $\alpha = \beta(=1)$, then the norm is called multiplicative on $O$, but only few algebras are norm–multiplicative.

Essentially, the implicit use of the set $O$ removes the redundancy in representing $T(X,Y)$, i.e. removing unnecessary direction pairs in $X \times Y$. Surely, the exact determination of the set $O$ is a combinatorial hard problem and depends on $T$ as well as on the subsets $X$ and $Y$. Moreover, this set in general lacks for linear or convex properties.

### A. Notation

In the following we will only consider $\mathbb{R}^N$ with standard inner product product and the corresponding Euclidean norm $\|\mathbf{x}\|^2 := \langle \mathbf{x}, \mathbf{x}\rangle$. For a given subset $X \subseteq \mathbb{R}^N$ we will denote the shell in $X$ with inner and outer radii $\alpha$ and $\beta$ by $X^{\alpha,\beta} := \{\mathbf{x} \in X \mid \alpha \leq \|\mathbf{x}\| \leq \beta\}$ and abbreviate further $X^\alpha := X^{0,\alpha}$. The functional $\|\mathbf{x}\|_0$ denotes the cardinality of the support of $\mathbf{x}$ in the Euclidean basis $\{\mathbf{e}_i\}_{i=0}^{N-1}$, i.e. the sparsity of $\mathbf{x}$ with respect to the Euclidean basis. A convex set $X \subset \mathbb{R}^N$ is a convex cone if for every $\mathbf{x} \in X$ and $\lambda \geq 0$ it follows $\lambda\mathbf{x} \in X$.

[1]Similar constructions are possible for $\mathbb{C}^N$.

$X$ has dimensionality $S$ if the space $\operatorname{span} X$ has dimension $S$.

### B. Main Theorem

Our main theorem provides a generalized compressed sensing framework by a stable embedding of certain $(S,F)$-sparse signal models which can not be anymore described by $K$-sparse signal models $\Sigma_K$. Since $T$ has the RNMP, there exists by Definition 1 a subset $O \subset X \times Y$ such that the representation by $T$ and $O$ leads to a stable embedding of channel outputs $T(\mathbf{s}, \mathbf{h})$ received over a fixed but unknown $F$-sparse channel state $\mathbf{h} \in Y$ if the difference set of all channel outputs $T(\mathbf{s}_1 - \mathbf{s}_2, \mathbf{h})$ is a subset of $T(X,Y)$.

*Theorem 1:* Let $2 \leq S, F, N, M \in \mathbb{N}$ with $SF \leq N$ and $X, Y \subset \mathbb{R}^N$ are $S$ resp. $F$ dimensional convex cones. If the bilinear map $T\colon X \times Y \to \mathbb{R}^N$ has the *restricted norm multiplicativity property* with bounds $\alpha$ and $\beta$, then a realization of a sub-Gaussian matrix $\Phi\colon \mathbb{R}^N \to \mathbb{R}^M$ with $M \leq N$ and $[\Phi]_{ij} \sim \mathcal{N}(0, 1/M)$ fulfills for every $\mathbf{z} \in T(X,Y)$

$$(1-\delta)\|\mathbf{z}\| \leq \|\Phi\mathbf{z}\| \leq (1+\delta)\|\mathbf{z}\| \tag{4}$$

for any $\delta \in (0,1)$ with probability

$$\geq 1 - 2N(X^1, X^{\delta/d})N(Y^1, Y^{\delta/d})e^{-c_0(\delta/2)M} \tag{5}$$

and constants

$$d = d(\alpha,\beta) := \begin{cases} 7\frac{\beta}{\alpha}(2+\sqrt{\alpha}) &, \alpha \neq \beta \\ 12 &, \alpha = \beta \end{cases} \tag{6}$$

$$c_0(\delta/2) := (3\delta^2 - \delta^3)/48. \tag{7}$$

*Remark 2:* Here $N(X^1, X^\epsilon) := \min\{n \mid \exists\{\mathbf{p}_i\}_{i=1}^n\colon X^1 \subset \bigcup_i (X^\epsilon + \mathbf{p}_i)\}$ denotes the covering number of $X^1$ by the covering sets $X^\epsilon$. The determination of the covering numbers is a Banach geometrical problem and well studied for various compact subsets of Banach-spaces [10].

*Remark 3:* To compute the probability for a mapping $\Phi$ which is universal for all $L = \binom{N}{S}\binom{N}{F} \leq N^{S+F}$ canonical $(S,F)$ dimensional convex cone pairs $(X,Y)$ we apply the union bound technique. Once universal RNMP bounds for all $L$ canonical cone pairs are found, we get the RIP with overwhelming probability for $M = \mathcal{O}((S+F)\log(N))$.

*Proof:* The main idea follows the technique in [2], where Baraniuk et al. considered a linear subspace $Z$ of $\mathbb{R}^N$ with a $\delta/4$-net $R$ for $Z^{1,1}$ and obtained by the measure concentration phenomenon of Gaussian matrices $\Phi$, that for every $\mathbf{r} \in R$ and any $\delta \in (0,1)$ it holds[2]:

$$|\|\Phi\mathbf{r}\| - \|\mathbf{r}\|| \leq \frac{\delta}{2}\|\mathbf{r}\| \tag{8}$$

with probability

$$> 1 - 2e^{-c_o(\delta/2)M}. \tag{9}$$

Both, the constant $\delta$ and the dimension of $Z$ determine then the cardinality of $R$, which is given by the covering number and yields the scaling of the exponential term in (9). This idea can be used again on $X$ and $Y$ as well to get an upper bound

[2]every $\mathbf{r} \in \mathbb{R}^N$ and the inequality (8) is equivalent to $\left|\|\Phi\mathbf{r}\|^2 - \|\mathbf{r}\|^2\right| \leq \delta/2\|\mathbf{r}\|^2$, see [2].

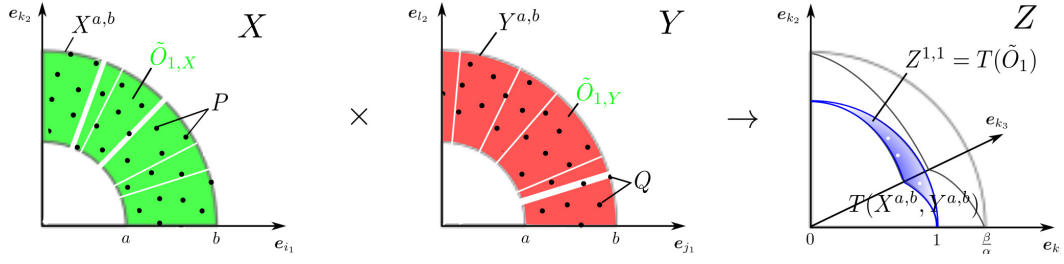

Figure 1. Net construction in the shells for covering the sphere in $Z$.

on the cardinality of $R$, but now in terms of the covering numbers $N(X^1, X^{\delta/d})$ and $N(Y^1, Y^{\delta/d})$. For this we need to control the norm of $\mathbf{z}$ by elements in $X \times Y$ which is possible if $T$ has the RNMP, since the set $O$ does not contain "bad" representation pairs for $Z := T(X, Y)$. It is in fact not necessary to give an explicit parametrization of $O$. The only information needed for the proof are the bounds $\alpha$ and $\beta$.

Every normalized $\mathbf{z} \in Z^{1,1}$ can be represented as an element from the image under $T$ of $O_1 := \{\mathbf{o} \in O \mid \|T(\mathbf{o})\| = 1\} \subset X \times Y$. Since $T$ has the RNMP on $O$ we have by Definition 1 for $(\mathbf{x}, \mathbf{y}) \in O_1$:

$$\alpha \|\mathbf{x}\| \|\mathbf{y}\| \leq 1 \leq \beta \|\mathbf{x}\| \|\mathbf{y}\|. \quad (10)$$

If we rescale the elements in the pair $(\mathbf{x}, \mathbf{y}) \in O_1$ by $\mu := \|\mathbf{x}\|$ resp. $\mu^{-1} > 0$ and set $\tilde{\mathbf{x}} := \mathbf{x}/\mu \in X^{1,1}$ and $\tilde{\mathbf{y}} := \mu \mathbf{y} \in Y$ ($X$ and $Y$ are cones), we get by bilinearity:

$$T(\tilde{\mathbf{x}}, \tilde{\mathbf{y}}) = \mathbf{z} = T(\mathbf{x}, \mathbf{y}) \quad (11)$$

and from absolute homogeneity of $\|\cdot\|$ we have with (10):

$$\alpha \|\tilde{\mathbf{y}}\| \leq 1 \leq \beta \|\tilde{\mathbf{y}}\| \quad \Rightarrow \quad \frac{1}{\beta} \leq \|\tilde{\mathbf{y}}\| \leq \frac{1}{\alpha}. \quad (12)$$

Hence there exist a representation set for $Z^{1,1}$ by pairs in $X^{1,1} \times Y^{1/\beta, 1/\alpha}$. Since $\alpha \leq 1 \leq \beta$ we can also choose a representation set $\tilde{O}_1$ which is contained in the symmetrized set of convex shells $X^{a,b} \times Y^{a,b}$ with common inner and outer radii $a := \beta^{-\frac{1}{2}} \leq 1 \leq \alpha^{-\frac{1}{2}} =: b$.

$$\tilde{O}_1 \subset \tilde{O}_{1,X} \times \tilde{O}_{1,Y} \subset X^{a,b} \times Y^{a,b}, \quad (13)$$

where $\tilde{O}_{1,X}$ and $\tilde{O}_{1,Y}$ are projections of $\tilde{O}_1$ to $X$ resp. $Y$. These are axial lines in the shells, see Fig. 1. Note, that $\tilde{O}_1$ can not be written as a Cartesian product, since only certain pairs are allowed.

The algebraic part of the proof goes as follows: Any realization of $\Phi$ is a linear map on a finite–dimensional normed space $\mathbb{R}^N$ and hence bounded, i.e. there exist $A \geq -1$ s.t.

$$\|\Phi \mathbf{z}\| \leq (1 + A) \|\mathbf{z}\| \quad \text{for all} \quad \mathbf{z} \in Z, \quad (14)$$

where $1 + A \geq 0$ denotes the smallest upper bound (operator norm of $\Phi$ restricted to $Z$). If we can show that $A \leq \delta$ we have shown the upper bound in (4). Since $\Phi$ is linear and $Z$ a cone (not necessarily convex), it is enough to find an upper bound in (14) for all $\mathbf{z} \in Z^{1,1}$. Let $P \subset X^{a,b}$ and $Q \subset Y^{a,b}$ be $\epsilon$-nets for $X^{a,b}$ resp. $Y^{a,b}$ with $\epsilon \in (0, 1)$ and define $R := \{T(\mathbf{p}, \mathbf{q}) : (\mathbf{p}, \mathbf{q}) \in P \times Q\} \subset Z$. It follows that $|R| \leq N(X^b, X^\epsilon) N(Y^b, Y^\epsilon)$. Thus, every $\mathbf{z} \in Z^{1,1}$ can now be represented with (13) by a pair $(\mathbf{x}, \mathbf{y}) \in X^{a,b} \times Y^{a,b}$ with

$x \in X^\epsilon(\mathbf{p}) := X^\epsilon + \mathbf{p}$ and $y \in Y^\epsilon(\mathbf{q}) := Y^\epsilon + \mathbf{q}$. The image $T(X^\epsilon(\mathbf{p}), Y^\epsilon(\mathbf{q}))$ is the covering set of the point $T(\mathbf{p}, \mathbf{q})$ and the union forms a covering for $Z^{1,1}$ by (13). Note that this covering sets in $Z$ are not necessarily convex! By using the triangle inequality and a zero addition in $\mathbf{p}, \mathbf{q}$ we have for any $T(\mathbf{p}, \mathbf{q}) \in R$ that all $\mathbf{z} = T(\mathbf{x}, \mathbf{y}) \in T(X^\epsilon(\mathbf{p}), Y^\epsilon(\mathbf{q})) \cap Z^{1,1}$ (i.e. $(\mathbf{x}, \mathbf{y}) \in \tilde{O}_1 \cap X^\epsilon(\mathbf{p}) \times Y^\epsilon(\mathbf{q})$) satisfy:

$$|\|\Phi \mathbf{z}\| - \|\Phi T(\mathbf{p}, \mathbf{q})\|| \leq \|\Phi(T(\mathbf{x} - \mathbf{p}, \mathbf{y} - \mathbf{q}))\|$$
$$+ \|\Phi(T(\mathbf{x} - \mathbf{p}, \mathbf{q}))\| + \|\Phi(T(\mathbf{p}, \mathbf{y} - \mathbf{q}))\|. \quad (15)$$

Using the universal bound $1 + A$ in (14) we obtain:

$$\leq (1 + A) \big( \|T(\mathbf{p}, \mathbf{y} - \mathbf{q})\| + \|T(\mathbf{x} - \mathbf{p}, \mathbf{q})\|$$
$$+ \|T(\mathbf{x} - \mathbf{p}, \mathbf{y} - \mathbf{q})\| \big) \quad (16)$$

and since $\mathbf{x} - \mathbf{p} \in X$ and $\mathbf{y} - \mathbf{q} \in Y$ we can apply the universal upper bound $\beta$ of the RNMP in (2) to get:

$$\leq (1 + A)\beta \big[ \|\mathbf{p}\| \|\mathbf{y} - \mathbf{q}\| + \|\mathbf{q}\| \|\mathbf{x} - \mathbf{p}\| + \|\mathbf{x} - \mathbf{p}\| \|\mathbf{y} - \mathbf{q}\| \big].$$

Since $(\mathbf{p}, \mathbf{q}) \in X^b \times Y^b$ is a $\epsilon$-net point-pair for $(\mathbf{x}, \mathbf{y})$, we get

$$\leq (1 + A)\beta \left( b\epsilon + b\epsilon + \epsilon^2 \right) \stackrel{\epsilon \leq 1}{\leq} (1 + A)\beta(2b + 1)\epsilon. \quad (17)$$

If we define the constant $c = c(\alpha, \beta)$ by:

$$c := \beta(2b + 1) = \beta \left( 2/\sqrt{\alpha} + 1 \right) > 1, \quad (18)$$

we obtain the upper bound

$$\|\Phi \mathbf{z}\| \leq (1 + A)c\epsilon + \|\Phi T(\mathbf{p}, \mathbf{q})\|. \quad (19)$$

The main tool of the proof is the measure concentration in (8). But there is no norm nesting $\|T(\mathbf{p}, \mathbf{q})\|$ since in general $(\mathbf{p}, \mathbf{q}) \notin \tilde{O}_1$. Even if $(\mathbf{p}, \mathbf{q}) \in \tilde{O}_1$ we don't get from (13) a tight scaling for vanishing $\epsilon$. Therefore we use the continuity property (bilinearity) of $T$ to upper and lower bound $\|T(\mathbf{p}, \mathbf{q})\|$ in terms of $\epsilon$ for every Cartesian product of two convex covering sets $X^\epsilon(\mathbf{p}), Y^\epsilon(\mathbf{q})$. Let us define the pre-image of $T(X^\epsilon(\mathbf{p}), Y^\epsilon(\mathbf{q})) \cap Z^{1,1}$ by

$$Z^{-1}(\mathbf{p}, \mathbf{q}) := \left\{ (\mathbf{x}, \mathbf{y}) \in X^\epsilon(\mathbf{p}) \times Y^\epsilon(\mathbf{q}) \cap \tilde{O}_1 \right\}.$$

If this set is not empty (otherwise $(\mathbf{p}, \mathbf{q})$ can be dropped from $P \times Q$), just grap one pair $(\mathbf{x}, \mathbf{y}) \in Z^{-1}(\mathbf{p}, \mathbf{q})$. But then there exist[3] $(\mathbf{c}, \mathbf{d}) \in X^\epsilon \times Y^\epsilon$ s.t. $(\mathbf{x}, \mathbf{y}) = (\mathbf{p} + \mathbf{c}, \mathbf{q} + \mathbf{d})$ and so:

$$\|T(\mathbf{p}, \mathbf{q})\| = \|T(\mathbf{x} - \mathbf{c}, \mathbf{y} - \mathbf{d})\|$$
$$= \|T(\mathbf{x}, \mathbf{y}) - T(\mathbf{x}, \mathbf{d}) - T(\mathbf{c}, \mathbf{y}) + T(\mathbf{c}, \mathbf{d})\|.$$

[3] If $X$ is a convex cone, then $\mathbf{p}$ is the aphex point of the covering set $X^\epsilon(\mathbf{p})$ which is again a convex cone (non-symmetrical), precisely $\epsilon X^1 = X^\epsilon$. Hence $\mathbf{x} - \mathbf{p} \in X^\epsilon$. If $X$ is a linear space, then $X^\epsilon$ is a ball (symmetric) with center at the origin and so $\mathbf{x} - \mathbf{p} \in X^\epsilon$ again.

Since $0 \leq \|\mathbf{c}\|, \|\mathbf{d}\| \leq \epsilon$ we get with (3) the lower bound

$$\|T(\mathbf{p},\mathbf{q})\| \geq \|T(\mathbf{x},\mathbf{y})\| - \|T(\mathbf{x},\mathbf{d})\| - \|T(\mathbf{c},\mathbf{y})\| - \|T(\mathbf{c},\mathbf{d})\|$$
$$\geq 1 - 2\beta b\epsilon - \beta\epsilon^2 \geq 1 - \beta(2b+1)\epsilon = 1 - c\epsilon$$

and the upper bound

$$\|T(\mathbf{p},\mathbf{q})\| \leq \|T(\mathbf{x},\mathbf{y})\| + \|T(\mathbf{x},\mathbf{d})\| + \|T(\mathbf{c},\mathbf{y})\| + \|T(\mathbf{c},\mathbf{d})\|$$
$$\leq 1 + 2\beta b\epsilon + \beta\epsilon^2 \leq 1 + \beta(2b+1)\epsilon = 1 + c\epsilon.$$

Let us discuss the discontinuity of this norm estimation. If we have $\alpha = \beta$, hence norm multiplicativity, then we would get $c = 3$. But in fact, this is to bad, since the shells are now unit spheres and every $\mathbf{p}, \mathbf{q}$ is normalized and hence by the norm multiplicativity $T(\mathbf{p},\mathbf{q})$. But this gives $c = 0$. To respect this fact we define $\tilde{c}$ and get for all net point pairs

$$1 - \tilde{c}\epsilon \leq \|T(\mathbf{p},\mathbf{q})\| \leq 1 + \tilde{c}\epsilon \quad , \quad \tilde{c} := \begin{cases} c & , \alpha \neq \beta \\ 0 & , \alpha = \beta \end{cases}. \quad (20)$$

Then we can use the measure concentration in (8) to obtain from (19) and (20) with probability larger than in (9)

$$\|\Phi\mathbf{z}\| \leq (1+A)c\epsilon + (1+\delta/2)(1+\tilde{c}\epsilon)$$
$$= 1 + Ac\epsilon + c\epsilon + \tilde{c}\epsilon + \delta(\tilde{c}\epsilon + 1)/2.$$

Now, by compactness, there exist a maximal $\mathbf{z}' \in Z^{1,1}$ such that equality in (14) is achieved. Hence we get

$$A \leq \frac{2c\epsilon + \tilde{c}\epsilon(2+\delta) + \delta}{2(1-c\epsilon)}. \quad (21)$$

Let us proceed by case distinction. If $\alpha = \beta$ then $\tilde{c} = 0, c = 3$. Defining $\epsilon = \frac{\delta}{12} \leq 1$ with $\delta \in (0,1)$ we get

$$A \leq \frac{3\epsilon + \frac{\delta}{2}}{1-3\epsilon} \leq \delta. \quad (22)$$

If we have $\alpha \neq \beta$ then $\tilde{c} = c = c(\alpha,\beta)$. Defining $\epsilon = \frac{\delta}{7c} \leq 1$ with $\delta \in (0,1)$ we get from (21):

$$A \leq \frac{\frac{5c\epsilon}{2} + \frac{\delta}{2}}{1-c\epsilon} \leq \frac{\frac{5\delta+7\delta}{14}}{1-\frac{\delta}{7}} \stackrel{\delta \leq 1}{\leq} \frac{\frac{12\delta}{14}}{\frac{6}{7}} = \delta. \quad (23)$$

This upper bound holds with probability larger than

$$> 1 - 2N(X^b, X^{\delta/\tilde{d}})N(Y^b, Y^{\delta/\tilde{d}})e^{-c_0(\delta/2)M},$$
$$\tilde{d} := \tilde{d}(\alpha,\beta) = \begin{cases} 7\beta(2/\sqrt{\alpha}+1) & , \alpha \neq \beta \\ 12 & , \alpha = \beta \end{cases}. \quad (24)$$

The lower bound $1 - \delta$ follows from this with

$$\|\Phi\mathbf{z}\| \geq \|\Phi T(\mathbf{p},\mathbf{q})\| - (1+A)c\epsilon \quad (25)$$

by considering all $\mathbf{z} \in Z^{1,1}$ we get by inserting (23) and (20) with same probability as in (24)

$$\|\Phi\mathbf{z}\| \geq \left(1 - \frac{\delta}{2}\right)\left(1 - \tilde{c}\frac{\delta}{\tilde{d}(\alpha,\beta)}\right) - (1+\delta)\frac{c\delta}{\tilde{d}(\alpha,\beta)}. \quad (26)$$

If $\alpha = \beta$ then $\tilde{c} = 0, c = 3$ and $\tilde{d} = 12$. This gives

$$\|\Phi\mathbf{z}\| \geq 1 - \delta/2 - \delta/2 = 1 - \delta. \quad (27)$$

If $\alpha \neq \beta$ then $\tilde{c} = c, \tilde{d} = 7c$ and

$$\|\Phi\mathbf{z}\| \geq 1 - \frac{\delta - c\epsilon\delta}{2} - c\epsilon - \frac{2c\delta}{7c} \geq 1 - \delta.$$

The covering number $N(X^b, X^{\delta/\tilde{d}})$ remains the same if we scale both sets $X^b, X^{\delta/\tilde{d}}$ by $1/b = \sqrt{\alpha}$, [10, Lemma 4.16] giving with $d := \tilde{d}/\sqrt{\alpha}$ the $\delta$-embedding with probability:

$$> 1 - 2N(X^1, X^{\frac{\delta}{d}})N(Y^1, Y^{\frac{\delta}{d}})e^{-c_0(\frac{\delta}{2})M}$$

∎

## III. APPLICATIONS

Before using the theorem we will discuss the following observations. A simple coupling $T$ in (1) is given by the *pointwise multiplication*, e.g. a fading channel:

$$T(\mathbf{s},\mathbf{h}) = \mathbf{H}\mathbf{s} = \mathbf{h} \odot \mathbf{s} := (h_0 s_0, \ldots, h_{N-1} s_{N-1})^T \quad (28)$$

with diagonal channel matrix $\mathbf{H} = \text{diag}(\mathbf{h})$. There, we have the norm inequality $\|\mathbf{h} \odot \mathbf{s}\| \leq \|\mathbf{h}\|\|\mathbf{s}\|$ and $\langle \mathbb{R}^N, +, \odot, \|\cdot\|\rangle$ becomes an *unital commutative algebra* with unit element $\mathbf{1}_\odot = (1, \ldots, 1)^T$. Unfortunately, one can not establish a lower norm multiplicativity bound $\alpha > 0$ on any disjoint convex subsets $X, Y \subset \mathbb{R}^N$. Hence we can not apply efficiently our theorem on these sets. But actually this is not necessary here, since for any $S, F$ dimensional subspaces $X, Y$ the output $\mathbf{z}$ has sparsity less than $\min\{S, F\}$ with respect to the Euclidean basis. In this simple case we can immediately apply the original Lemma 5.1 in [2] to establish the RIP on $X \odot Y$ with probability

$$> 1 - 2(12/\delta)^{\min\{S,F\}} e^{-c_0(\delta/2)M}. \quad (29)$$

This easily extents to pointwise multiplications in another domains, i.e. for some given unitary matrix $\mathbf{U}$ we can define the new product (commutative as well):

$$T(\mathbf{s},\mathbf{h}) = \sqrt{N}\mathbf{U}^*(\mathbf{U}\mathbf{s} \odot \mathbf{U}\mathbf{h}), \quad (30)$$

which obeys the following $\ell^2$-norm inequality:

$$\frac{1}{N}\|T(\mathbf{s},\mathbf{h})\|^2 = \|\mathbf{U}^*(\mathbf{U}\mathbf{s} \odot \mathbf{U}\mathbf{h})\|^2 = \sum_i |\langle\mathbf{e}_i,\mathbf{U}\mathbf{s}\rangle\langle\mathbf{U}\mathbf{h},\mathbf{e}_i\rangle|^2$$
$$\leq \max_j |\langle\mathbf{e}_j,\mathbf{U}\mathbf{s}\rangle|^2 \|\mathbf{h}\|^2 \leq \max_{i,j} |\langle\mathbf{e}_j,\mathbf{U}\mathbf{e}_i\rangle|^2 \|\mathbf{s}\|_1^2 \|\mathbf{h}\|^2$$
$$\leq \|\mathbf{U}\|_\infty^2 \|\mathbf{s}\|_0 \|\mathbf{s}\|^2 \|\mathbf{h}\|^2$$

with $\|\mathbf{U}\|_\infty := \max_{i,j}|\langle\mathbf{e}_j,\mathbf{U}\mathbf{e}_i\rangle|$ and $\|\mathbf{s}\|_1 := \sum_j |\langle\mathbf{s},\mathbf{e}_j\rangle|$. Hence, by commutativity of $T$, we obtain for any unitary matrix $\mathbf{U}$ the upper estimate:

$$\|T(\mathbf{s},\mathbf{h})\|^2 \leq N \|\mathbf{U}\|_\infty^2 \min\{\|\mathbf{s}\|_0, \|\mathbf{h}\|_0\} \|\mathbf{s}\|^2 \|\mathbf{h}\|^2. \quad (31)$$

*a) Sparse Circular Convolutions:* Let us consider multiplication in the frequency domain, i.e. $\mathbf{U} = \mathbf{F}$ is the Fourier-matrix with $[\mathbf{F}]_{lk} = e^{-2\pi i \frac{lk}{N}}/\sqrt{N}$ for $l, k \in \{0, \ldots, N-1\}$. Then $T = \circledast$ is the *circular convolution* in the time domain by (30). With $\|\mathbf{F}\|_\infty = 1/\sqrt{N}$ we obtain from (31)

$$\|\mathbf{h} \circledast \mathbf{s}\|^2 \leq \min\{\|\mathbf{h}\|_0, \|\mathbf{s}\|_0\} \|\mathbf{h}\|^2 \|\mathbf{s}\|^2. \quad (32)$$

If $\mathbf{s}$ and $\mathbf{h}$ are sparse in the Fourier basis, we can proceed as before. But for $(S,F)$–sparsity in the time domain (with respect to the Euclidean basis) we need RNMP for $\circledast$. We call such a restricted circular convolution an $(S,F)$-sparse circular convolution, see [1]. In general there doesn't exists a lower bound $\alpha > 0$ for sparse circular convolutions, since the nullspace of $T$ can contain elements $(\mathbf{s},\mathbf{h})$ with $\mathbf{s} \neq \mathbf{0}$ *and* $\mathbf{h} \neq \mathbf{0}$. One can prevent this behavior by restricting the domain of $T$ to certain convex sets $X, Y \subset \mathbb{R}^N$.

*b) Sparse Circular Convolutions on Positive Cones:*
If we assume the channel is only an on/off channel or a fading channel with positive parameters $h_i \geq 0$ and the signal parameters are also positive, then we can easily establish the following lower bound:

$$\|\mathbf{h} \circledast \mathbf{s}\|^2 = \sum_{k,j} h_j^2 s_k^2 + \underbrace{\sum_{k,j,l \neq j} h_j h_l s_{k\ominus j} s_{k\ominus l}}_{\geq 0} \geq \|\mathbf{h}\|^2 \|\mathbf{s}\|^2.$$

Hence we obtain together with (32) for positive inputs

$$\|\mathbf{h}\| \|\mathbf{s}\| \leq \|\mathbf{h} \circledast \mathbf{s}\| \leq \sqrt{\min\{S,F\}} \|\mathbf{h}\| \|\mathbf{s}\|. \quad (33)$$

The positive elements in $\operatorname{span}\{\mathbf{e}_i\}_{i \in I}$ with $I \subset \{0, \ldots, N-1\}$ form a canonical (Euclidean) $S = |I|$-dimensional positive convex cone $X$. If $X$ and $Y$ are canonical $S$ resp. $F$ dimensional positive convex cones, then we can apply our theorem with the norm bounds derived in (33), to establish the RIP on $X \circledast Y$. With a result of Rogers-Zhong [11] and Rogers [12] we can find an upper bound for the covering number of $S \geq 3$ dimensional positive cones $X^1$ by $N(X^1, X^\epsilon) \leq (4/\epsilon)^S 7S \log S$. For $S = 2$ we can upper bound this by the covering number with $S$-dim $\tilde{\epsilon}$-balls contained in $X^\epsilon$, where $\tilde{\epsilon} = \epsilon/(2\sqrt{2})$. Hence $(3/\tilde{\epsilon})^2 = (6\sqrt{2}/\epsilon)^2$ $\tilde{\epsilon}$-balls resp. $\epsilon$-cones cover the whole unit ball and hence $X^1$. But this number is less than $(4/\epsilon)^2 14 \log 2$ and thus Rogers bound holds also for all $S \geq 2$. A rough estimate shows that Rogers bound can be upper bounded for $S \geq 2$ by $(18/\epsilon)^S$. Since $\alpha = 1$ and $\beta = \sqrt{\min\{S,F\}}$ in (33) we get $d = \tilde{d} = 21\beta$ with (24). For $\epsilon = \delta/d$ we can hence establish the RIP on $X \circledast Y$ by Theorem 1 with probability

$$\geq 1 - 2\left(378\sqrt{\min\{S,F\}}/\delta\right)^{S+F} e^{-c_0(\delta/2)M}. \quad (34)$$

*c) Sparse Circular Convolutions and Tensor Products:*
A main property of the circular convolution is that the image of Euclidean basis vectors $(\mathbf{e}_i, \mathbf{e}_j)$ is again an Euclidean basis vector. Let $X := \operatorname{span}\{\mathbf{e}_i\}_{i \in I}$ and $Y := \operatorname{span}\{\mathbf{e}_j\}_{j \in J}$ with $|I| = S$ and $|J| = F$. Then it is easy to show that $X \circledast Y = \operatorname{span}\{\mathbf{e}_k\}_{k \in I \oplus J}$ where $I \oplus J := \{(i+j) \mod N \mid i \in I, j \in J\}$. If $|I \oplus J| = SF$ then the subspaces $X, Y$ are "properly" separated [1] and one can show easily that the image is Hilbert isomorph (isometric) to the set of all simple tensor products. This means, there exist isomorphism, s.t. on these extremal pairs $(X, Y)$ the norm $\|\cdot\|$ becomes *multiplicatively*. Hence we get for all $(\mathbf{s}, \mathbf{h}) \in X \times Y$

$$\|\mathbf{s} \circledast \mathbf{h}\| = \|\mathbf{s}\| \|\mathbf{h}\|. \quad (35)$$

Here we use the fact, that every simple tensor $\tilde{\mathbf{s}} \otimes \tilde{\mathbf{h}}$ has a unique representation up to scalars by $\tilde{\mathbf{s}}$ and $\tilde{\mathbf{h}}$, [13]. This implies a norm multiplicativity, i.e. $\|\tilde{\mathbf{s}} \otimes \tilde{\mathbf{h}}\| = \|\tilde{\mathbf{s}}\| \|\tilde{\mathbf{h}}\|$.

The covering number for an $S$ dimensional ball $X^1 \subset \mathbb{R}^N$ with $X^\epsilon$ can be upper bounded by $N(X^1, X^\epsilon) \leq (3/\epsilon)^S$ [10]. Together with the norm multiplicativity (35), $1 = \alpha = \beta$, we get $d = 12$ and obtain by Theorem 1 as probability bound

$$> 1 - 2(36/\delta)^{S+F} e^{-c_0(\delta/2)M}. \quad (36)$$

## IV. Conclusions

In many communication schemes the coupling of channel- and signal parameters is given by bilinear maps $T$ and sparsity is present in both inputs to $T$. It is therefore of general interest for several engineering applications to characterize the compressibility of the whole output set. In this paper we provide such a characterization once the RNMP can be established, uniformly or in some probabilistic setting. However, a uniform treatment of the exemplary case of $(S,F)$–sparse circular convolutions is still an open problem. It is also not fully understood whether the RNMP is only a sufficient or really a necessary condition for a scaling with $\mathcal{O}(S+F)$. In particular, it is important to know whether to the following conjecture is true: Let $(X,Y)$ be $(S,F)$–sparse canonical subspaces in $\mathbb{R}^N$ $(S, F \geq 2)$. Does then any realization of a sub-Gaussian matrix $\Phi: \mathbb{R}^N \to \mathbb{R}^M$ with $M \leq N$ and $[\Phi]_{ij} \sim \mathcal{N}(0, 1/M)$ fulfills for every $\mathbf{z} \in X \circledast Y$

$$(1-\delta)\|\mathbf{z}\|^2 \leq \|\Phi\mathbf{z}\|^2 \leq (1+\delta)\|\mathbf{z}\|^2 \quad (37)$$

with failure probability $p_e \leq 2(d/\delta)^{S+F} e^{-c_0(\delta/2)M}$ for some fixed constant $d > 1$ and every $\delta \in (0,1)$?

### Acknowledgment

The authors would like to thank Holger Boche and David Gross for their helpful discussion on this topic. This work was partly supported by the Deutsche Forschungsgemeinschaft (DFG) grants Bo 1734/13-1 and JU 2795/1-1&2.